\documentclass[final,3p,twocolumn]{elsarticle}
\usepackage{amssymb}
\usepackage{graphicx}
\usepackage{lineno}

\journal{Nuclear Instruments and Methods A}

\begin{document}

\begin{frontmatter}

\title{Resonant feedback for axion and hidden sector dark matter searches}

\author{Edward.\,J.\,Daw}
\address{Department of Physics and Astronomy, The University of Sheffield, Hicks Building, Hounsfield Road, Sheffield S3 7RH, United Kingdom}

\date{\today}

\begin{abstract}
Resonant feedback circuits are proposed as an alternative to normal modes of conducting wall cavities or lumped circuits in searches for hidden sector particles. The proposed method offers several potential advantages over the most sensitive axion searches to date, including coverage of a wider range of axion masses, the ability to probe many axion masses simultaneously, and the elimination of experimentally troublesome mechanical tuning rod mechanisms. After an outline of the proposed method, we present a noise budget for a straw-man experiment configuration. We show that the proposed experiment has the potential to probe the axion mass range 
$\rm{2-40\,\mu eV}$ with 38 days of integration time. Other existing and proposed resonant searches for hidden sector particles may also benefit from this approach to detection. 
\end{abstract}


\end{frontmatter}

\section{Introduction}
\label{sec:introduction}

The identification of dark matter remains a central unsolved problem in astrophysics \cite{bergstrom:2012fi}. Dark matter candidates beyond the standard model include the QCD axion \cite{Weinberg:1977ma,Wilczek:1977pj}, which could be a significant component of dark matter and solve the strong CP problem \cite{Peccei:1989a,Peccei:2006as}, as well as other hidden sector fields \cite{Feng:2008ya,Ahlers:2007qf}. In a tuned resonant cavity halo axion detector, first proposed by Sikivie \cite{Sikivie:1983ip}, axions from our local halo undergo Primakoff conversion \cite{Pirmakoff:1951pj} in a static magnetic field $\vec{B}$, converting to real photons at a rate enhanced by a factor of $Q$, the quality factor of a mode of a normal-conducting walled cavity \cite{jackson:1975a}.

Those cavity axion detectors which have published search results
\mbox{[11-15]}
have consisted of copper plated cylindrical cavities of unloaded $Q\sim10^5$. The cavity volume encloses one or more moveable tuning rods used to control the $\rm{TM}$ mode frequencies. The $\rm{TM_{010}}$ mode has the largest form factor for photons converting from dark matter axions thermalized in the galactic halo. The energy of such photons is given by $h\nu=m_ac^2[1+\mathcal{O}(v^2/c^2)]$, where $\nu$ is the electromagnetic mode frequency, $v$ is a velocity typical of halo axions, and the kinetic energy term is of order $10^{-6}$ of the axion rest energy. Recently, the ADMX experiment has achieved sensitivity to the DFSZ \cite{Dine:1981rt,Zhitnitsky:1980tq} axion model for the first time \cite{Du:2018uak}, using microwave SQUID amplifiers (MSAs) to achieve detector noise temperatures at the $\rm{150\,mK}$ level \cite{Mueck1,Mueck2}.

Other hypothesized lightly coupled particles having a variety of spins and parities may be dark matter constituents. Such so-called hidden sector particles may play multiple significant roles in particle physics beyond the standard model, astrophysics, and cosmology. There have been many experiments and new proposals to probe a range of energy scales for new hidden sector fields 
[24-40].
In many of these experiments, resonant detection techniques, based on the same principles as the axion search described above, are used to enhance the rate of conversion to detectable electromagnetic signals. The necessity of tuning these detectors, dwelling at each tuning for sufficient time to achieve an acceptable signal to noise ratio, is a problematic limiting factor for all such resonant designs.

In this paper, we propose exploitation of a resonant feedback technique as an alternative to cavities, lumped circuits or other resonant structures. Feedback techniques have been considered previously by Rybka in the context of increasing the quality factor for a cavity resonance \cite{Rybka:2014ria}, but here we propose feedback as a technique to induce entirely new resonances suitable for axion searches. There are two components to the proposed scheme, the first being an electromagnetic structure with a high form factor for hidden sector fields to convert to electromagnetic fields, but not itself possessing a resonance at the frequency of interest. The second component is a filter circuit whose transfer function replicates that of a high $Q$ resonator at the desired frequency. These two elements are connected together in a closed feedback loop, with open loop gain and phase shift just below the Nyquist stability limit \cite{Nyquist:1932a}. We demonstrate that this detector has potential to yield similar resonant enhancement of the rate for axion conversion into photons as a conducting wall resonant cavity with the same resonant frequency and quality factor. We suggest that the same structures could be used in resonant searches for other hidden sector particles. We show further that suitable resonant filters can be realised using digital signal processing elements controlled parametrically through the filter coefficients. We derive expressions for the closed loop gain and quality factor of the system, the condition for stability of the circuit, and the quality factor of the resonance. We calculate a noise budget for an example configuration, showing that the performance requirements of each element shared with conventional designs are no more stringent than is the case in conventional receivers, and that the addition of room temperature digital electronics does not significantly increase the noise temperature of the apparatus.

The proposed technique offers three significant advantages over standard resonant configurations. Firstly, parametric control of the mode parameters removes the need for cryogenic mechanical or electronic tuning elements such as tuning rods in resonant cavities. Secondly, the operation of many such digital filters in parallel means that the apparatus may cover many possible hidden sector particle masses in parallel. Thirdly, the explicit link between the length scale of the electromagnetic structure used as the detector and the wavelength of the conversion photons is broken, offering the possibility of searching a greater range of particle using the same magnet bore.

\section{Proposed Detection Scheme}
\label{sec:scheme}

In Sikivie's resonant axion detector scheme, the axion field interacts with a static magnetic field, resulting in an oscillating electromagnetic field that excites a mode of an electromagnetic resonator.
The conversion signal power is proportional to a form factor given by
\begin{equation}
C=\frac{\left|\int_V\vec{E}\cdot\vec{B}dV\right|^2}
{\int_V\varepsilon\left|\vec{E}\right|^2dV\int_V\left|\vec{B}\right|^2dV},
\label{eq:formfactor}
\end{equation}
where $\vec{E}$ is the amplitude of the oscillating electric field from axion conversion. This form factor is integrated over the physical geometry of the electromagnetic configuration into which the axions convert. In the cavity detection scheme, the form factor is evaluated over the volume of the cavity, and the electric field is the position-dependent electric field of the mode used to sense axions. For cylindrical cavities containing off-center tuning rods, the form factor is evaluated from fields calculated using finite element modeling. Readout of cavity detectors is accomplished by coupling the cavity mode to an external amplifier using an electric field probe. Matching of the field probe to the external amplifier is accomplished by varying the insertion depth of the probe until the scattering parameter of the reflected signal off the probe, $\rm{s_{11}}$, is at the desired coupling level. The resonant frequency is determined by the positions and geometries of the conducting walls of the cavity and the tuning rods for the mode in question.

\begin{figure}[h!]
\begin{center}
\includegraphics[width=0.8\columnwidth]{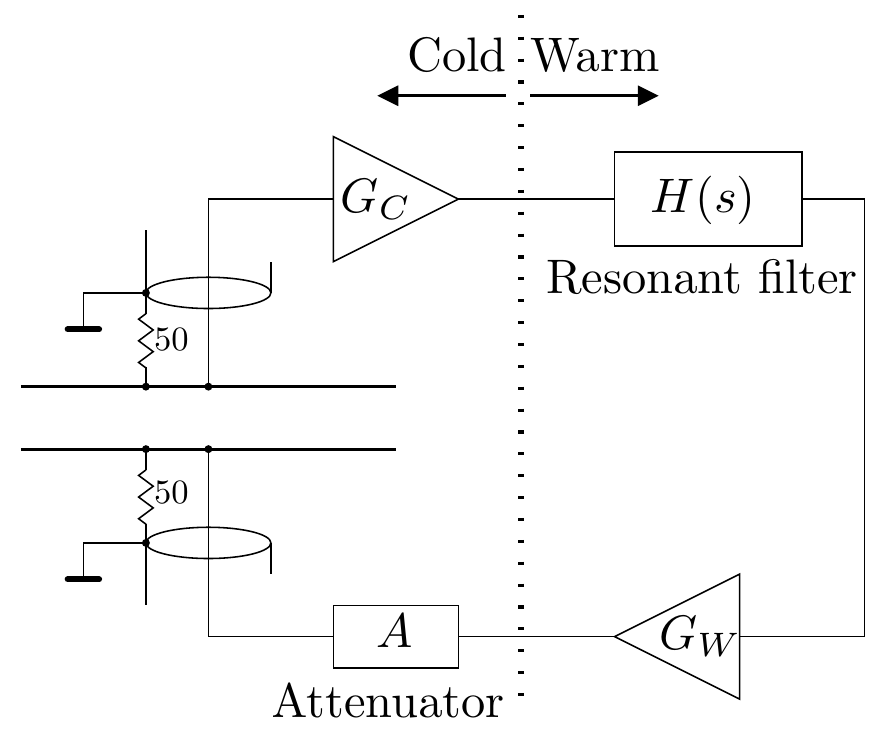}
\end{center}
\caption{\label{fig:schematic1} A schematic of the concept for a modified Sikivie-type axion detector utilizing resonant feedback. The resonant mode of the cavity is replaced by an external resonant circuit, and the cavity itself is replaced by a parallel plate capacitor coupled to the external resonator by transmission lines connected to the capacitor plates.}
\end{figure}

\begin{figure}[h!]
\begin{center}
\includegraphics[width=0.9\columnwidth]{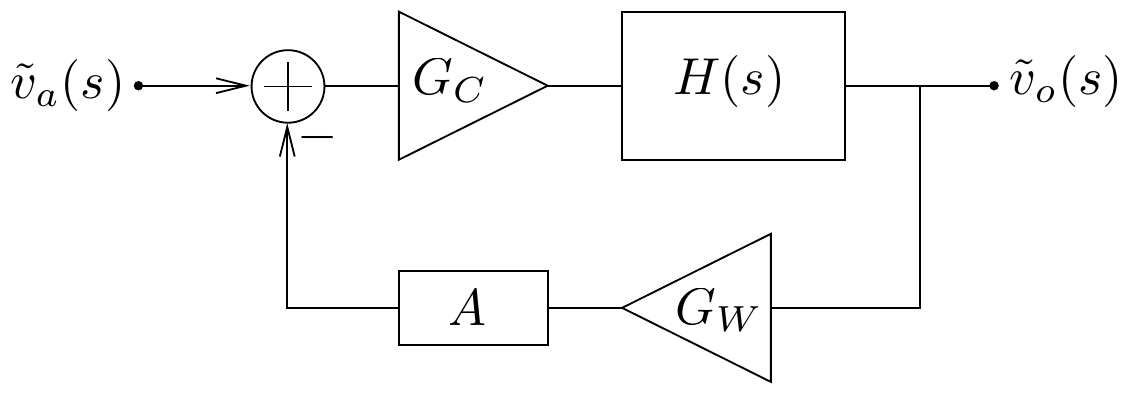}
\end{center}
\caption{\label{fig:equivcircuit} An equivalent circuit representation of the closed loop feedback circuit, showing the coupling of an axion signal into the circuit and the key elements of the resonant system, including cryogenic and warm amplifiers and a cryogenic attenuator.}
\end{figure}

Figure \ref{fig:schematic1} shows the proposed alternative scheme, in which axions convert within a parallel plate capacitor operated far below cutoff, with the applied magnetic field normal to the capacitor plates. The capacitor plates are coupled to transmission lines connecting them to an external circuit, with those capacitor plates matched to the resistive $\rm{50\,\Omega}$ characteristic impedance of the transmission lines by discrete fixed resistors connecting the capacitor plates to the transmission line ground. At frequencies around $\rm{1\,GHz}$, the magnitude of the reactance of a parallel circular plate capacitor having $\rm{1.5\,cm}$ plate separation and $\rm{50\,cm}$ diameter is
$\rm{1.4\,\Omega}$, so the impedance mismatch due to the capacitor is not significant.
The resonant filter, discussed further in the Appendix 
has transfer function
\begin{equation}
H(s)=\frac{\Gamma s}{s^2+\Gamma s + \omega_0^2},
\label{eq:openlooptf}
\end{equation}
where $s=i\omega$ and $\omega=2\pi f$, where $f$ is the frequency, $\Gamma/(2\pi)$ is the frequency full width at half maximum of the resonance, and $\omega_0/(2\pi)$ is the resonant frequency of the filter. Figure \ref{fig:equivcircuit} shows an equivalent circuit relating the output signal to the axion-induced signal on the plate of the parallel-plate capacitor.
The closed loop transfer function of this circuit is
\begin{equation}
H_C(s)=\frac{\tilde{v}_o(s)}{\tilde{v}_i(s)}=\frac{G_C\Gamma s}{s^2+(1+AG_CG_W)\Gamma s+\omega_0^2}.
\label{eq:closedlooptf}
\end{equation}
Comparing with the open loop transfer function of Equation \ref{eq:openlooptf}, we see that the width of the resonant peak is modified compared to the open loop case by a factor of $1+AG_CG_W$. As for any feedback circuit, raising the gains too high will drive the loop  into oscillation \cite{Nyquist:1932a}. The open loop round trip transfer function of the feedback circuit at the resonant frequency is
\begin{equation}
H_0(i\omega_0)=AG_CG_W,
\label{eq:openlooptfomega0}
\end{equation}
which is less than unity as long as the attenuation of the signal at the fixed attenuator is sufficient to render the round trip gain to be less than 1. We shall see in Section \ref{sec:noisebudget} that a large attenuation is also necessary to ensure that the noise budget in the readout circuit gives sufficient signal to noise ratio of the axion signal against noise contributions from the cold and warm readout components.

\begin{figure}
\begin{center}
\includegraphics[width=0.8\columnwidth]{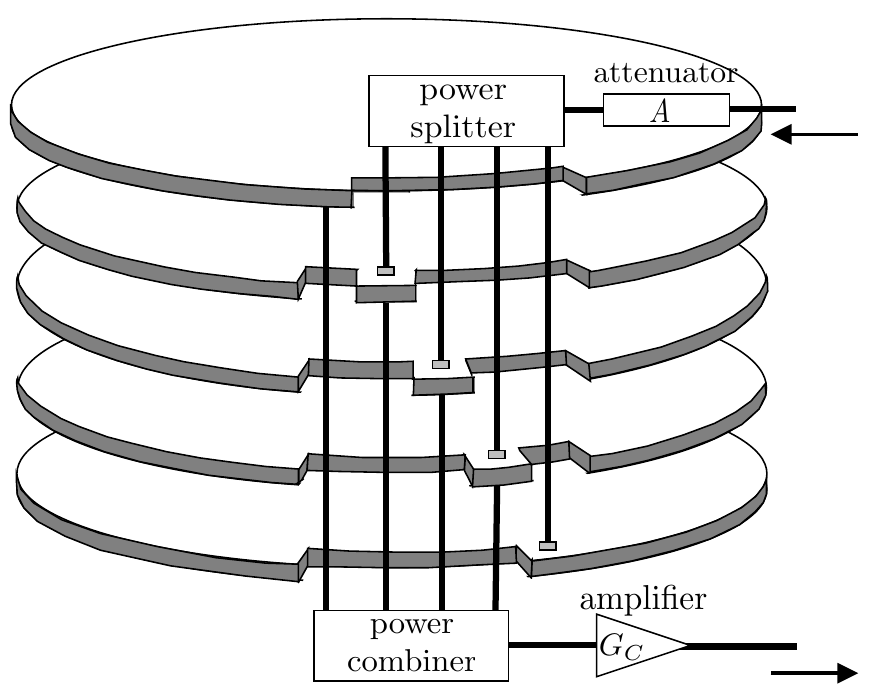}
\end{center}
\caption{\label{fig:capstack} Four capacitative axion sensors in a stack, with the upper (lower) plates of the vacuum-gap capacitors power combined (split), in-phase, with a Wilkinson type power combiner (splitter). Each circular disk is a circuit board copper clad on each face. All cables are coaxial. The cold attenuator and low noise cold amplifier also shown in Figures \ref{fig:schematic1}, \ref{fig:equivcircuit}, \ref{fig:noiseloop} and \ref{fig:digitalcavity} are included. The arrows indicate the direction of signal propagation around this portion of the resonant feedback circuit. The cables between the capacitor plates and the power combiners must all have the same electrical length. This could be accomplished either by cable loops on the connections to the closer plates or by fabrication of a custom Wilkinson combiner and splitter on circuit boards above and below the stack. The electrical connections between the capacitor plates and the coaxial cables to the Wilkinson devices each use the scheme shown in Figure \ref{fig:schematic1}. A stack of diameter
$\rm{25\,cm}$ and of plate separation $\rm{1.5\,cm}$ having 5 plates as shown would have height $\rm{\sim6\,cm}$. Increasing the number of capacitors to 64 whilst maintaining the same capacitor geometry would result in a 220 litre volume. A small stack such as that shown could be run in parallel with a cavity in an existing axion experiment as a commissioning test and proof of principle.}
\end{figure}

In practice, the capacitor plates would be separated by insulating spacers and small microwave substrate circuit boards mounted on the back of the plates would be used to attach the connectors to coaxial transmission lines and the resistive terminations. The volume of a magnet bore could be filled with a stack of capacitors connected in parallel, thereby decreasing the reactive impedance presented to the electronics by a factor of the number of capacitors. Figure \ref{fig:capstack} is an illustration of such a parallel capacitor stack; 4 parallel capacitors combined in-phase are shown, but the idea could be expanded to, for example, 64 capacitors each of diameter $\rm{25\,cm}$ and plate separation 
$\rm{1.5\,cm}$ yielding a total stack height of 1m, the same geometry as the current ADMX cavity. In this schematic the electrical connections are close to the edges of the capacitor plates, however, connections could also be made close to the plate centers by feeding the cables through holes in the intermediate plates. Finite element modelling studies will determine the most appropriate scheme.

Dark matter axions, having de Broglie wavelengths of hundreds of metres produce a conversion signal coherent across the capacitor array, and in each capacitor the resulting oscillating electric field is parallel to the applied magnetic field inside the magnet bore. Transmission lines to the capacitors would be matched in physical length to ensure that the signals from axion conversion at the outputs of transmission lines from each capacitor in the array, and these transmission lines would be terminated on Wilkinson passive power combiners, the single output of which would be fed to the cryogenic amplifier, with the same arrangement in reverse leading from the cold attenuator output back to each of the capacitors. In this arrangement the form factor for axion to photon conversion is close to unity. The transmission lines connected to individual capacitor plates would be combined in-phase using Wilkinson power combiners; therefore the impedance mismatch of $\rm{1.4i\,\Omega}$ due to the capacitative load does not get larger as more capacitors are combined.

\section{Resonant feedback compared to resonant cavities}
\label{sec:capcavcomp}
\begin{figure}
\begin{center}
\includegraphics[width=1.0\columnwidth]{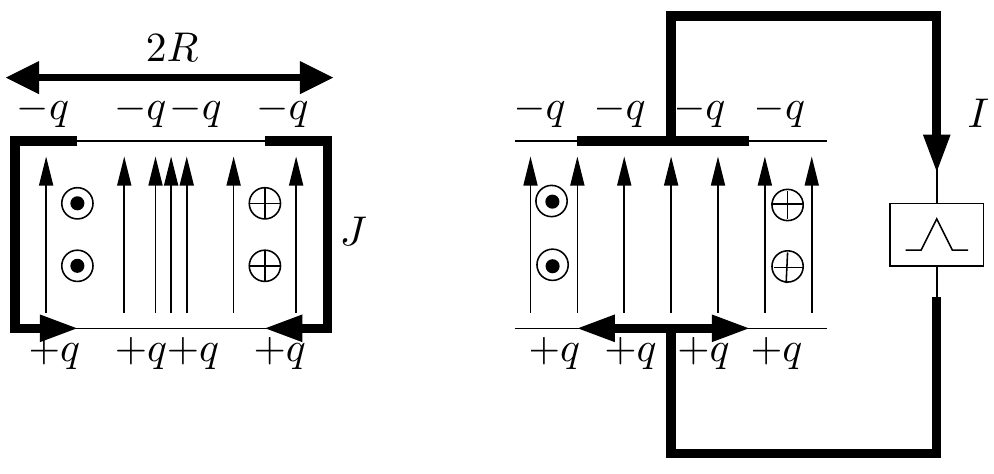}
\end{center}
\caption{\label{fig:capcavcomp} On the left, a cross section through a right circular cylindrical conducting wall resonant cavity excited in the $TM_{010}$ mode. The vertical arrows indicate the electric field lines, and the circles with dots (crosses) indicate magnetic field lines directed out of (in to) the plane of the diagram. Half a period later the direction of both electric and magnetic field lines reverses. The electric field lines terminate on charge distributed on the inner wall of the metal ends. Currents flowing in the side walls, the charges on the ends, and the time rate of change of the displacement current, $\varepsilon_0\partial \vec{E}/\partial t$ completes the electrical circuit. In the case of the capacitor, the electric field is more uniform (we have exaggerated the distance between the capacitor plates relative to the plate size for clarity), and the circuit is completed by the external wire that passes through the resonant electronics.} 
\end{figure}

It is not obvious that the circuit with the external resonant filter achieves the same resonant enhancement as the Sikivie cavity technique, In this section we show semi-quantitatively that the proposed resonant feedback scheme should achieve similar enhancement in signal power by a factor of $Q$, previously demonstrated both experimentally and theoretically for the cavity case in both the classical and single quantum limits \cite{Sikivie:1983ip, Kleppner:1989, Jaeckel:2011a}. Figure \ref{fig:capcavcomp} shows a cross section through a right circular cylindrical metal wall resonant cavity with the $TM_{010}$ mode excited on the left, and a capacitor of similar geometry connected to a simplified resonant feedback circuit on the right.
In the cavity case, the high Q resonance is due to boundary conditions on $\vec{E}$ and $\vec{B}$ at the cavity walls. As with all resonators, fields circulating in the cavity form closed circuits, with power building up when the phase shift in the round trip is such that constructive interference occurs. In the case of the $TM_{010}$ mode, the magnitude of the electric field is given as a function of time $t$ and the cylindrical radial coordinate $\rho$ as
\cite{jackson:1975a}
\begin{equation}
E(\rho,t)=E_0J_0\left(\frac{2.405\rho}{R}\right)\cos(\omega t).
\label{eq:besselfield}
\end{equation}
These standing waves may be decomposed into travelling waves whose argument circulates about the origin in the complex plane, but whose magnitude drops off as the waves travel radially using Hankel functions \cite{Abramowitz},
\begin{equation}
\begin{array}{rcl}
E(\rho,t)&=&\frac{E_0}{2}\Re\left\{\left[H_0^{(1)}\left(\frac{2.405\rho}{R}\right)\right.\right.+\\
&&\left.\left.H_0^{(2)}\left(\frac{2.405\rho}{R}\right)\right]e^{-i\omega t}\right\}.\\
\end{array}
\label{eq:hankeldecomp}
\end{equation}
The Hankel functions $H_0^{(1)}$ and $H_0^{(2)}$ are complex valued, rotating, respectively, anticlockwise and clockwise in the complex plane with increasing $\rho$. Since both terms have the same sign of the time dependent phasor $e^{-i\omega t}$, we see that these terms represent complex fields of amplitude that drops off with increasing $\rho$, propagating inwards and outwards along the cylindrical radial direction, respectively. These waves make round trips from the wall to the centre and then back again to the wall, where they interfere with themselves yielding constructive interference at a frequency of
$\nu_{010}=2.405c/R$, which for $\mathrm{R=0.25m}$ is $\rm{459\,MHz}$. Axions convert to photons that excite the $TM_{010}$ cavity mode when the electric field of the mode is parallel to an externally applied magnetic field because axions modify Maxwells equations, contributing an extra displacement current proportional to 
$\vec{B}\partial a/\partial t$  and hence exciting this mode.
Figure \ref{fig:equivcircuit} shows a representation of axions adding a signal to just such a circulating field. In \cite{Jaeckel:2011a} a similar circulating field analysis is performed for an optical resonator used as a detector in a light shining through a wall experiment. The critical condition for the axion to excite the resonance is that it remain coherent over many round-trip times of the circulating fields. In the case of the $TM_{010}$ mode, the round trip circulation time for this field is 
$t_c=2R/c$, which is $\rm{1.7\,ns}$ when $R=\rm{25\,cm}$.

An estimate of the coherence time of the dark matter halo axion field is the ratio of the De Broglie wavelength of halo axions to a typical velocity. Our bare cavity resonant frequency corresponds to an axion mass of 
$\mathrm{1.9\mu eV}$. With a halo dark matter virial velocity of $\beta c=\rm{240\,km\,s^{-1}}$, the De Broglie 
wavelength of the axion field is $\lambda=(2\pi\hbar c)/(\beta mc^2)$, or $\rm{826\,m}$. Therefore the coherence
time of the axion field is $\tau_{\rm{coh}}=h/(\beta^2mc^2)=\mathrm{3.4\,ms}$, the axion field is coherent over 2 million cycles of the circulating fields underlying the cavity mode.

In the resonant feedback circuit, the circulating current travels around the external circuit instead of down the cavity walls. The signal path passes out of the cryostat through room temperature receiver electronics, and back through an RF feed through attenuators and back to the other capacitor plate. We assume that the feedback path has a physical length of 20m and consists mostly of RG401 cable, having a signal propagation velocity of $\rm{0.7c}$, An additional delay is induced by the ADC and DAC, which sample at 250MHz, and delay the signal by two sampling periods, or 8ns. The total time delay round the loop is therefore $\tau_d=\rm{103\,ns}$. The ratio of the axion coherence time
$\tau_{\rm{coh}}$ to this delay time is $N_d=\tau_{\rm{coh}}/\tau_d=34,000$. This ratio limits the quality factor of the resonance in the loop that can be induced by axions having this coherence time. It can be interpreted as an
effective quality factor of the resonant enhancement,$Q_d=\pi N_d$, corresponding to an effective $Q_d$ of 107,000. This $Q_d$ is greater than that of the loaded cavity resonance in resonant cavity detectors. Hence, the enhancement of the signal power due to the presence of the feedback loop should be comparable with that of the $TM_{010}$ mode of a cavity. Improvements in $Q_d$ are possible mainly by shortening the loop, which is certainly possible but requires careful detector design as the digital signal processing electronics is unlikely to work below 
a temperature of $\rm{243\,K}$, may fail in a high magnetic field, and probably emits digital noise which precludes close proximity to the capacitor or superconducting electronics.

\section{Noise Budget}
\label{sec:noisebudget}
\begin{figure}[h!]
\begin{center}
\includegraphics[width=1.0\columnwidth]{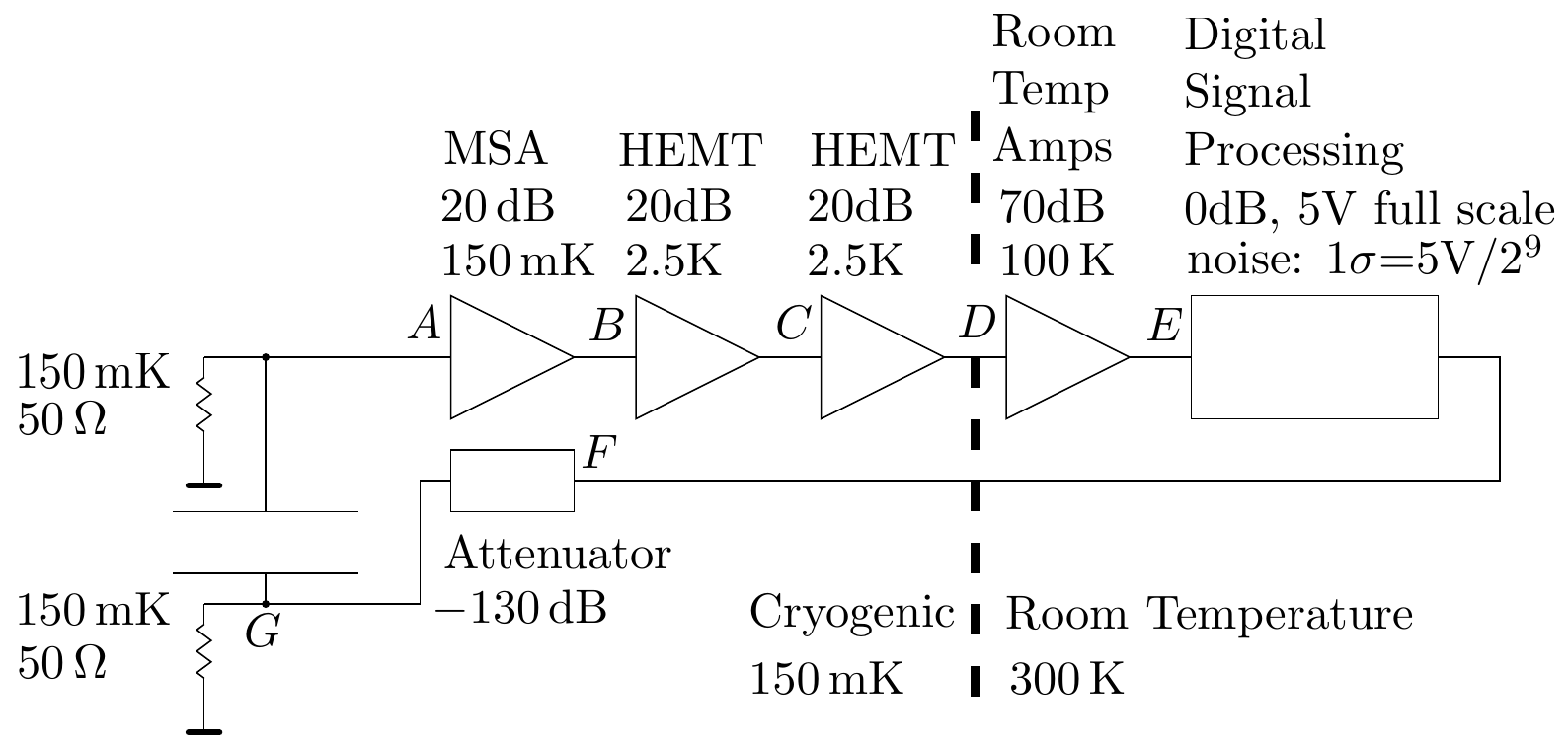}
\end{center}
\caption{\label{fig:noiseloop} A schematic of the practical implementation of feedback electronics, showing ultra-low-noise receiver electronics common to this proposal and more conventional cavity searches, as well as the feedback path to and from the capacitor, operated far below cut-off, threaded by a large static magnetic field. MSA and HEMT are acronyms for `microwave squid amplifier' and `high electron mobility transistor', respectively. The noise performance of MSAs, critical to these searches, are discussed in \cite{Mueck1, Mueck2}. HEMT amplifiers for ultra high frequency, low noise applications are discussed in \cite{HEMTs1, HEMTs2}.}
\end{figure}

The proposed feedback circuit incorporates both low noise cryogenic electronics and noisier room temperature digital electronics. In this section, we show that using existing amplifier technology it is possible nevertheless to maintain a good signal-to-noise ratio for mode oscillations around the feedback loop. Figure \ref{fig:noiseloop} shows a more realistic arrangement of amplification stages as implemented in a representative axion search. This noise budget is approximate; in practice the a detailed noise model of the electronics must be constructed to understand the spectral content of the noise background, especially in the vicinity of sharp features such as high Q resonances.
The temperature $T$ of a source of Johnson noise in a bandwidth $B$ emits noise power $P$ into a balanced sink given by
\begin{equation}
P=k_BTB.
\label{eq:johnsonnoise}
\end{equation}
At a mode frequency of $\rm{700\,MHz}$ corresponding to an axion mass of $\rm{2.9\,\mu eV}$, assuming a magnetic field of $\rm{6.8\,T}$, and other experimental parameters corresponding to the ADMX2 experiment configuration \cite{Du:2018uak}, the signal power expected from a KSVZ \cite{Kim:1979if,Shifman:1979if} model axion converting to photons is $\rm{2\times10^{-22}\,W}$ in a bandwidth of $\rm{750\,Hz}$. This signal power, as well as the noise power in the same bandwidth may be calculated at different points around the feedback loop, labeled A-F. 

The signal power at the input to the digital electronics is -60dBm, corresponding to $\rm{220\,\mu V}$ across a $\rm{50\,\Omega}$ load. Assume that the digitiser has a full scale input voltage of 5V, a sampling rate of $\rm{250\,MHz}$, and 12 bits of precision with a $\rm{1\,\sigma}$ voltage noise at the 3rd least significant bit, the voltage noise is $\rm{25\,\mu V}$, so that the noise spectral density is $\rm{0.9\,\mu V/\sqrt{Hz}}$, and the noise contributed by the ADC in a $\rm{750\,Hz}$ bandwidth again referred to a $\rm{50\,\Omega}$ load is $\rm{-79\,dBm}$. Assuming the same level of noise from the DAC, the noise contribution of the ADC plus DAC is $\rm{-76\,dBm}$. The digital electronics therefore makes a negligible contribution to the overall noise, which is dominated by the contributions from the physical temperature of the source resistor and the first stage cryogenic amplifier. 

The attenuator must be cooled to as low a physical temperature as possible, probably by placing it in thermal contact with the refrigerator cold plate. If the temperature of the noise incident on the attenuator is $T_I$, the attenuation factor (ratio of output power to input power for matched source and sink) is $a$, and the attenuator temperature is $T_A$, then the noise temperature at the attenuator output is
\begin{equation}
T_O=aT_I+(1-a)T_A.
\label{eq:attenuation}
\end{equation}
Using Equations \ref{eq:johnsonnoise} and \ref{eq:attenuation}, and taking the values for component noise temperatures and power gains assumed in Figure \ref{fig:noiseloop}, we obtain the noise and signal powers at different points in the loop given in Table \ref{tab:noisebudget}.
\begin{table}
\caption{\label{tab:noisebudget} Noise budget around the closed loop resonant circuit shown schematically in Figure 3. The second column is total Johnson noise power into a 750Hz bandwidth, including both noise from the output of the previous stage, and noise contributed by the noise temperature at the input of the next component after the labelled location. The third column is noise power into the same bandwidth including only that due to the next component after the labelled location. The fourth column is the signal power. The attenuator will be set so that the open loop gain is in fact slightly less than 0dB to avoid the circuit going into oscillation, but here the difference between the actual open loop gain and 0dB is neglected.}
\begin{tabular}{| p{0.8cm} | p{1.4cm} | p{2.8cm} | p{1.0cm} |}
\hline
Point 
& 
Noise in 750Hz bandwidth 
& 
Noise from local component into 750Hz bandwidth
&
Signal power 
\\
& [dBm] & [dBm] & [dBm] \\
\hline \hline
A & -175 & -178 & -190 \\
B & -155 & -166 & -170 \\
C & -135 & -166 & -150 \\
D & -115 & -150 & -130 \\
E & -45 & -76 & -60 \\
F & -45 & -178 & -60 \\
G & -175 & -178 & -190  \\
\hline
\end{tabular}
\end{table}
The noise budget illustrates the fact that the digital electronics used to implement the resonant gain stage makes a negligible contribution to the noise budget in the cryogenic portion of the circuit, since this noise is attenuated by the large (roughly $\rm{130\,dB}$) attenuation of the cryogenically cooled attenuator that brings the open loop gain back to slightly less than 0dB to suppress spontaneous oscillation of the feedback circuit.

The signal to noise ratio around the loop is $\rm{-15\,dB}$, which is a power ratio of $1/32$. Integration is used to detect the power excess due to axions against the background of fluctuations about the average Johnson noise at surrounding frequencies, where the signal to noise ratio, $\rm{SNR}$, is given by the radiometer equation,
\begin{equation}
\rm{SNR}=\frac{P_S}{P_N}\sqrt{Bt},
\label{eq:radiometer}
\end{equation}
where $B$ is the $\rm{750\,Hz}$ signal bandwidth, $t$ is the integration time, and $P_S$ and $P_N$ are the power in the axion signal and the noise power in a $\rm{750\,Hz}$ bandwidth, respectively. Detection of KSVZ \cite{Kim:1979if,Shifman:1979if} axions at a signal to noise ratio of 4 requires an integration time of $\rm{21.8\,s}$. DFSZ axions \cite{Dine:1981rt,Zhitnitsky:1980tq} have a smaller signal power by a factor of $(0.36/0.98)^2=0.14$, reducing the ratio of $P_S$ to $P_N$ to $0.0044$, so that an integration time of $\rm{1120\,s}$  is required to achieve an SNR of 4 for DFSZ axions.

\section{Covering the mass range}
\label{sec:masscoverage}
We now calculate the integration time to search for halo DFSZ axions in the mass range $\rm{2-40\,\mu eV}$ using the proposed detector. A possible additional advantage of the proposed experiment is that the form factor for the electric fields of parallel plate capacitors operated far below cutoff is unity, compared to the $0.4$ typical of cavity modes. This boosts the signal power by a factor of $2.5$. However, there may be volume losses because of instrumentation between the capacitors in the stack that reduces this form factor, so we have not enhanced the signal power by this factor of 2.5 in the following calculation.

The key to speeding up a search for axions using the proposed instrument is that the digital resonant filters can be built up in parallel sets, since they are realized as digital filters on field programmable gate arrays as discussed in the Appendix. It is reasonable to assume that we can implement $100$ such resonators in parallel. There is easily enough space on a large FPGA for $100$ parallel resonant circuits. The circuits must have resonances separated by many resonance widths, and with each resonance approximately $\rm{14\,kHz}$ wide, allowing for 10 full widths separation between resonators, 100 resonators occupy $\rm{14\,MHz}$ of bandwidth, well matched to the bandwidth of a single MSA (RF Squid amplifier) \cite{Mueck1,Mueck2}. Intermediate masses between the resonances are probed by shifting all the resonant frequencies together in small steps so that uniform coverage is achieved.

We can therefore calculate the total integration time to cover the mass range $\rm{2-40\,\mu eV}$ . We can assume that the form factor, volume, and coupling of the axions to photons remain static. We assume 100 parallel resonators, with each resonator covering a bandwidth of approximately $\rm{15\,kHz}$ per scan. The integration time per scan is $\rm{1120\,s}$ to achieve DFSZ sensitivity. Therefore in $\rm{1120\,s}$ we cover a frequency range of $\rm{1.5\,MHz}$. There is no degradation in the anticipated signal power with increasing frequency, and we assume the noise temperature of the electronics can be maintained at a level of roughly $\rm{150\,mK}$. A target axion mass search range of
$\rm{2-40\,\mu eV}$, corresponds to a range of mode frequencies of $\rm{480\,MHz-4.8\,GHz}$, a bandwidth of
$\rm{4.34\,GHz}$, which is a factor of $2896$ times the $\rm{1.5\,MHz}$ per $\rm{1120\,s}$ integration needed for DFSZ sensitivity. Therefore the total search time for the targeted axion mass range is $\rm{1120\times2896\,s}$ or $37.5$ days. This experiment places a very large range of axion masses at the DFSZ model signal strength within reach with achievable integration times, the speed-up compared to that in previous searches being due to the availablity of many high Q, high form factor resonators in parallel.

\section{Conclusions}
\label{sec:conclusions}
We have presented a concept for a resonant feedback variant on the Sikivie resonant axion detector. We have shown how such resonant feedback could be used with simple electromagnetic structures that can be stacked up to fill the bore of a high field magnet, and how such structures can be impedance matched to low noise electronics of the type used by previous and current axion searches. We have studied the noise contribution of the newly proposed digital electronics, and concluded that it does not adversely affect the signal-to-noise ratio in the chain. We have shown that with 100 parallel resonant filters, the detection rate for axions is sufficiently enhanced that a range of axion masses previously only just accessible to Sikivie-type detectors may be probed in tens of days rather than years. We very much hope that the proposed scheme can play a part in the detection of axions within a reasonable experiment lifetime.
Other experimental groups have proposed resonant search experiments for axions and hidden sector particles proposing to utilise resonantly enhanced detection; see for example
[27, 29, 30-35, 38-40].

There are several areas in which further research connected to this idea is necessary. Though we believe the analysis presented here demonstrates that in the classical limit this arrangement will yield a sensitive axion detector, it is important to analyse this circuit in the quantum limit, since what we have essentially proposed here is a harmonic oscillator distributed between a `cold' domain in which the excitation of the oscillator consists of a small number of
quanta, and a `warm' domain in which the same level of excitation is represented as a much amplified signal in the classical limit. Further work will be necessary to understand how this rather interesting variety of oscillator works in detail.

Experimental tests of this idea are being planned. In a first test, a closed loop system could be built at room temperature and tested by using a network analyser coupled through a pair of field probes to the space between the capacitor places. The transfer function of such an arrangement in transmission should show the resonances generated by the digital filter, and by cooling the capacitor plates, it should be possible to see the resonant peaks in the Johnson noise spectrum of the electromagnetic radiation between the plates using a suitable low noise amplifier and spectrum analyser.

Further testing could be carried out in existing cavity axion search experiments. An unused RF port can be used as an injection port for the feedback signal, and the existing cryogenic amplifier chain can be used as far as the output of the last amplifier before the mixer. A parallel set of receiver electronics implementing external resonance can be used to generate an artificial resonance far from the physical resonances of the cavity as a demonstration that the generation of such resonances by feedback is feasible.

Finally, the idea of filling the magnet bore with a stack of capacitors needs to be tested in finite element modelling, to ensure that the capacitor electric fields behave as expected, and to work out where the transmission lines couple to the capacitor plates. In particular it would be very convenient to couple the transmission lines to the plates at the edges, near to the wall of the magnet bore, rather than in the middle, since the other capacitors in the stack would then be an obstruction. It may also be necessary to have more than one path to each capacitor plate, driven in phase, and this idea can also be studied using finite element modelling tools. Finally, note that existing axion search experiments may have unused space inside their cryogenic inserts that could be utilised for proof-of-principle tests at low temperature and in high magnetic fields in parallel with a more conventional tuned cavity search.

\appendix
\section{Digital resonant filter electronics}
\label{sec:digitalfilter}
\begin{figure}[h!]
\begin{center}
\includegraphics[width=\columnwidth]{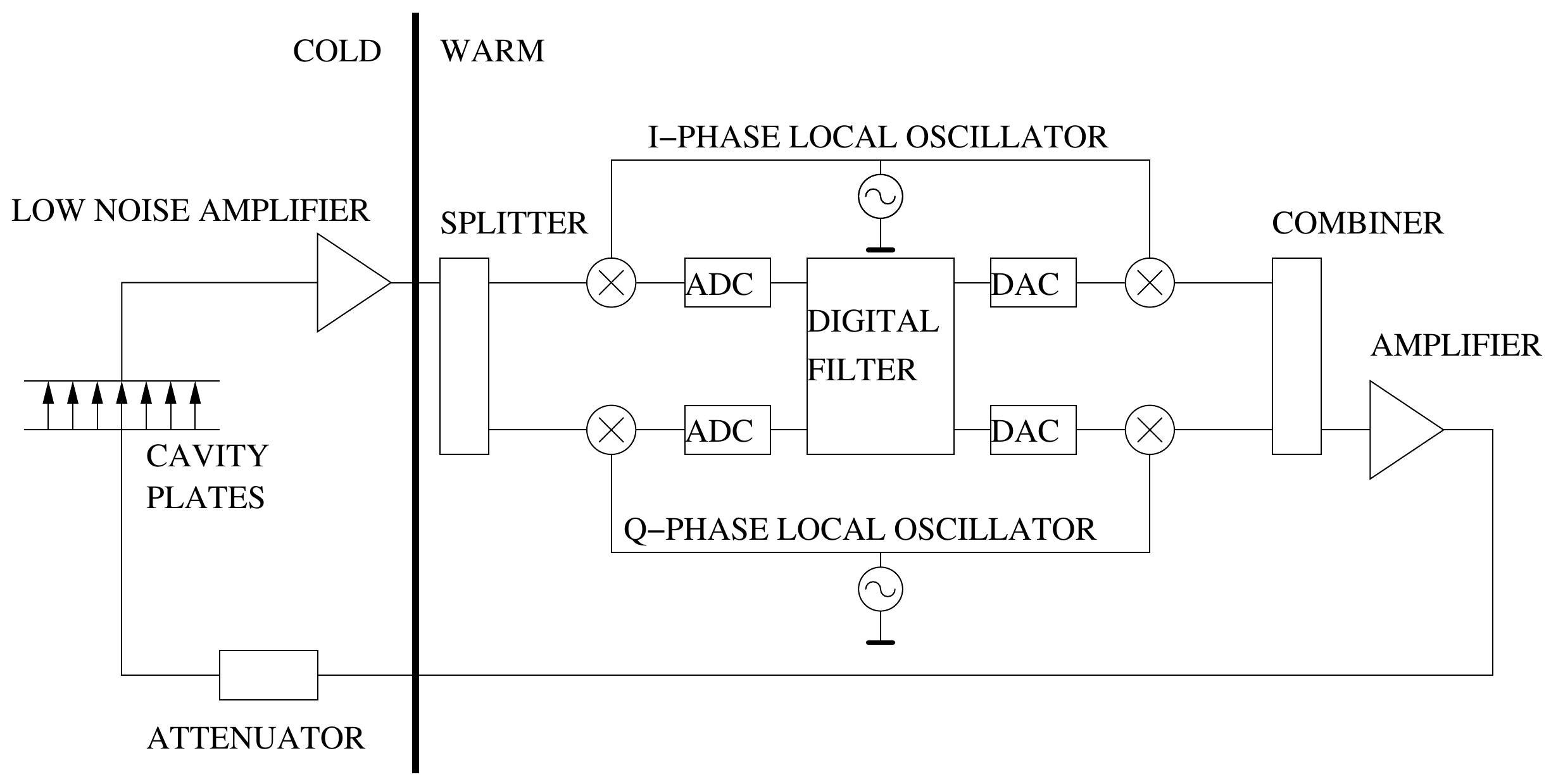}
\end{center}
\caption{\label{fig:digitalcavity} A schematic of the practical implementation of the digital portion of the electronics. The RF signal is mixed down in two orthogonal quadradratures using image-reject mixers before a digital filter implements the resonant circuit. The room temperature amplification stages following the cryogenic low noise amplifier and preceding the splitter are omitted here.}
\end{figure}

The resonant filter alluded to throughout this paper is here described in some detail. The filter is implemented digitally using a platform that allows continuous causal processing of the data. Field programmable gate arrays (FPGAs) providing a suitable architecture. The filter is an infinite impulse response type, and is fed by a stream of data, one sample at a time, the data being digitised from the output of the amplifier chain. There are several options for the practical aspects of this digitisation. In recent cavity axion searches, the output of the amplifier chain is heterodyned to a lower frequency using an image reject mixer, so that digitisation occurs with a sampling rate around $\rm{10\,MHz}$. The proposed idea could utilise this technique, but would utilise a higher digitisation rate, around $\rm{64\,MHz}$, and use two image-reject mixers in parallel, so that two out-of-phase IF quadratures are obtained from the RF signal stream. Both quadratures are digitised and the resulting data is passed through the resonant filter. Figure \ref{fig:digitalcavity} shows a schematic of the digital electronics. In reality, a single local oscillator would be used, fed through a $90^{\circ}$ hybrid to yield two out-of-phase local oscillator signals, each of which would be common to the corresponding pair of mixers before and after the digital electronics.
The resonant filter itself utilises an iteration algorithm originally developed for the monitoring of sinusoidal backgrounds in gravitational wave detectors. If we represent the two quadratures of the n\textsuperscript{th} sample of input data as the real and imaginary parts of a complex number $x_n$, and similarly represent the n\textsuperscript{th} sample of filter output as the real and imaginary parts of a second complex number $y_n$, then the filter algorithm takes the following form,
\begin{equation}
y_n=e^{-w+i\Delta_0}y_{n-1}+(1-e^{-w})x_n.
\label{eq:iwave}
\end{equation}
The steady state response of the filter to a phasor input, $x_n=e^{in\Delta}$, where $\Delta$ is a frequency in radians per sample, is $y_n=A(\Delta)e^{i\left(n\Delta+\phi(\Delta)\right)}$. Here $A(\Delta)$ and $\phi(\Delta)$ are the magnitude and phase of this transfer function, respectively. Substituting $x_n$ and $y_n$ into Equation \ref{eq:iwave}, we obtain the following expressions for $A(\Delta)$ and $\phi(\Delta)$:
\begin{equation}
\begin{array}{rcl}
A(\Delta)&=&\frac{1-e^{-w}}{\sqrt{1-2e^{-w}\cos(\Delta-\Delta_0)+e^{-2w}}}\\
\phi(\Delta)&=&\mathrm{atan}\left(\frac{e^{-w}\sin(\Delta-\Delta_0)}{1-e^{-w}\cos(\Delta-\Delta_0)}\right)\\
\end{array}
\label{eq:tf1}
\end{equation}
In the limit where $w\ll 1$ and $\delta=\Delta-\Delta_0$ much smaller than
$\Delta_0$, the magnitude of the transfer function becomes
\begin{equation}
A(\delta)\simeq\frac{1}{\sqrt{1+\frac{\delta^2}{w^2}}}.
\label{eq:tf2}
\end{equation}
This Lorentzian lineshape, characteristic of all high-Q damped oscillators, can also be obtained from the transfer function of Equation \ref{eq:openlooptf} by taking 
\begin{equation}
\left|H(\delta=\omega-\omega_0)\right|=\left.\sqrt{H(s)H(-s)}\right|_{\omega^2=-s^2}
\label{eq:tf3}
\end{equation}
in the limit where both $\Gamma$ and $\delta$ are much smaller than $\omega_0$.
The angular resonant frequency of the filter of Equation \ref{eq:iwave} is $\omega_0=\Delta_0/\tau_s$, where the sampling rate is $f_s=1/\tau_s$. The full width at half maximum (FWHM) is $\Gamma=2w/\tau_s$, and the quality factor is $Q=\Delta_0/(2w)$. The characteristic response time of the filter is $2\pi\tau_s/w$ seconds. This demonstrates that the action of the digital filter is that of a high Q resonant circuit, and justifies the use of Equation \ref{eq:openlooptf} to represent the resonant digital filter in Section \ref{sec:scheme}.

The filter can be implemented using only fixed point arithmetic such that the delay per sample is a single sampling period. The clock rate of $\rm{250\,MHz}$  common on modern FPGAs, with faster models also available, makes these calculations tractable at the proposed sampling rate, which provides adequate bandwidth to support 100 parallel resonances separated from each other by 10 full-widths in frequency space. The factor of $e^{i\Delta}$ means that the sine and cosine of $\Delta$ must be calculated for each desired resonant frequency in each parallel filter; these values could be stored in memory connected to the FGPA or calculated in parallel with filtering in preparation for the next frequency shift using a processor core on the FPGA alongside the filter circuits. Further speed-up in the calculations can be achieved by requiring that $w=2^{-N}$ with integer $N$, in which case the multiplications by factors of $w$ can be implemented with $N$ bit-shifts to the right, avoiding some of the fixed point multiplies. There is the potential for the FPGA device to also carry out other aspects of the data analysis such as the Fourier transforms necessary as the first stage in the search for axions in the data.

\section*{Acknowledgments}
This work was funded in part under the U.K. Science and Technology Facilities Council (STFC) grant ST/N000277/1, and partly through a grant provided by the University of Sheffield strategic development fund. The author wishes also to thank the ADMX collaboration, and in particular Prof. Leslie Rosenberg, Prof. Gray Rybka and Dr. Gianpaolo Carosi for useful and productive discussions, Mr. Mitchell Perry for work on early hardware prototypes of the digital filter, and Prof. Clive Speake, Dr. Ian Bailey, Prof. John March-Russell, and Prof. Sir Keith Burnett for helpful discussions of science and practical details related to this idea.


\end{document}